\documentclass[aip,apl,preprint]{revtex4}
\usepackage{graphicx}
\usepackage{epsfig}
\usepackage{color}
\usepackage{bm}

\begin{document}
\draft
\title{Resonant Scattering of Surface Plasmon Polaritons by Dressed Quantum Dots}

\author{Danhong Huang$^{1}$, Michelle Easter$^{2}$, Godfrey Gumbs$^{3}$,\\
A. A. Maradudin$^{4}$, Shawn-Yu Lin$^{5}$, Dave Cardimona$^{1}$,\\
and Xiang Zhang$^{6}$}
\affiliation{$^{1}$Air Force Research Laboratory, Space Vehicles
Directorate, Kirtland Air Force Base, NM 87117, USA\\
$^{2}$Department of Mechanical Engineering, Stevens Institute of Technology, 1 Castle Point Terrace, Hoboken, NJ 07030, USA\\
$^{3}$Department of Physics and Astronomy, Hunter College of the City University of New York, 695 Park Avenue New York, NY 10065, USA\\
$^{4}$Department of Physics and Astronomy and Institute for Surface and Interface Science,
University of California, Irvine, CA 92697, USA\\
$^{5}$Department of Physics, Applied Physics, and Astronomy,\\
Department of Electrical, Computer, and Systems Engineering, and Future Chips Constellation,\\
Rensselaer Polytechnic Institute, 110 8th Street, Troy, New York 12180, USA\\
$^{6}$National Science Foundation Nanoscale Science and Engineering Center,\\
3112 Etcheverry Hall, University of California at Berkeley, Berkeley, CA 94720}

\date{\today}

\begin{abstract}
The resonant scattering of surface plasmon-polariton waves by embedded semiconductor quantum dots
above the dielectric/metal interface is explored in the strong-coupling regime.
In contrast to non-resonant scattering by a localized dielectric surface defect,
a strong resonant peak in the scattering field spectrum is predicted and accompanied
by two side valleys. The peak height depends nonlinearly on the amplitude of
surface plasmon-polariton waves, reflecting the feedback dynamics from a photon-dressed
electron-hole plasma inside the quantum dots. This unique behavior in the scattering field
peak strength is correlated with the occurrence of a resonant dip in the absorption spectrum of
surface plasmon-polariton waves due to interband photon-dressing effect.
Our result on the scattering of surface plasmon-polariton waves may be experimentally observable and
applied spatially selective illumination and imaging of individual molecules.
\end{abstract}
\pacs{PACS:}
\maketitle

Most of the previously reported  research carried out on the optical properties of
condensed matter, including well-known optical absorption and inelastic
light scattering, made use of a weak probe field as a perturbation to the
system under investigation\,\cite{r1}. In this weak-coupling limit, the optical response
of electrons depends only on the material characteristics. However, with increased
field intensity, the optical properties of materials are found to depend nonlinearly on
the strength of the external perturbation\,\cite{r2}.
\medskip

Strong photon-electron interaction in semiconductors is known to produce dressed
states\,\cite{r20,r21,r22} with a Rabi gap for electrons and create substantial
nonlinearity in the semiconductor\,\cite{r3,r4,r5}.
The presence of an induced polarization field, treated as a source term\,\cite{r6} arising
from photo-excited electrons in metals, allows for a {\em resonant} scattering of surface
plasmon-polariton waves\,\cite{r7}. This is quite different from the {\em non-resonant} scattering
of surface plasmon-polariton waves by a localized dielectric surface defect\,\cite{r8,r9} or by
surface roughness\,\cite{r12}.
\medskip

In this Letter, based on the obtained analytical solution\,\cite{r10,r11} for the Green's function
of the coupled quantum dot and semi-infinite metallic material system, we
report for the first time our semi-analytic solutions of the self-consistent
equations for strongly coupled electromagnetic field dynamics and quantum kinetics
of electrons in a quantum dot above the surface of a thick metallic film. In our formalism,
strong light-electron interaction is manifested in the photon-dressed electronic states
and in the feedback from induced optical polarization of dressed electrons to the
incident light as well. Our calculated results predict a strong resonant peak in the
scattering field spectrum, which is accompanied by two side valleys at the same time.
Furthermore, we have discovered that the peak height varies nonlinearly with the amplitude
of the surface plasmon-polariton waves. This clearly demonstrates the effect due to
feedback dynamics from photon-dressed electron-hole (e-h) plasma inside quantum dots.
In addition, this unique observation in the scattering field spectrum is proven to be
correlated with a resonant dip observed in the absorption spectrum\,\cite{r7} of surface plasmon-polariton
waves which can be directly attributed to an effect of the inter-band photon-dressing of electronic states.
\medskip

Our model system consists of a semi-infinite metallic material and a semiconductor
quantum dot above its surface. A surface plasmon-polariton (SPP) wave is locally
excited through a surface grating by a normally-incident light. This propagating SPP wave
further excites an inter-band e-h plasma within the quantum dot. The induced local
optical polarization field of the photo-excited e-h plasma is resonantly coupled to
the SPP wave to produce a splitting in degenerate e-h plasma and SPP modes with an anti-crossing gap.
\medskip

By using the Green's function ${\cal G}_{\mu\nu}({\bf r},{\bf r}^\prime;\,\omega)$,
we may express Maxwell's equation for the electric field component
$\mbox{\boldmath$E$}({\bf r};\,\omega)$ of an electromagnetic field
in a semi-infinite non-magnetic medium in   position-frequency space as a
three-dimensional integral equation, i.e.,

\begin{equation}
E_{\mu}({\bf r};\,\omega)=E^{(0)}_{\mu}({\bf r};\,\omega)-
\frac{\omega^2}{\epsilon_0c^2}\sum\limits_{\nu}\int d^3{\bf r}^\prime\,
{\cal G}_{\mu\nu}({\bf r},{\bf r}^\prime;\,\omega)\,
{\cal P}_{\nu}^{\rm loc}({\bf r}^\prime;\,\omega)\ ,
\label{e4}
\end{equation}
where $\mbox{\boldmath$E$}^{(0)}({\bf r};\,\omega)$ is a solution in the
absence of semiconductor quantum dots, ${\bf r}=(x_1,x_2,x_3)$ is a three-dimensional
position vector, $\omega$ is the angular frequency of the incident light,
$\epsilon_0$  and $c$ are the permittivity and speed of light in vacuum,
$\mbox{\boldmath${\cal P}$}^{\rm loc}({\bf r};\,\omega)$ is an off-surface
local polarization field that is generated by optical transitions of electrons
in a quantum dot, which generally depends on the electric field in a nonlinear
way and should be determined by the optical Bloch equations. Additionally, the
position-dependent dielectric constant $\epsilon_{\rm b}(x_3;\,\omega)$
is equal to $\epsilon_{\rm d}$ for the semi-infinite dielectric material
in the region  $x_3>0$, but is given by $\epsilon_{\rm M}(\omega)$
for the semi-infinite metallic material in the region $x_3<0$.
\medskip

By assuming a surface plasmon-polariton wave propagating within the $x_1-x_2$-plane,
we can write

\begin{equation}
\mbox{\boldmath$E$}^{(0)}({\bf r};\,\omega_{\rm sp})=E_{\rm sp}\,e^{i{
\bf k}_0(\omega_{\rm sp})\cdot{\bf D}_0}\,\frac{c}{\omega_{\rm sp}}
\left[i\hat{\bf k}_0\beta_3(k_0,\omega_{\rm sp})-\hat{\bf x}_3k_0(\omega_{\rm sp})\right]
\,e^{i{\bf k}_0(\omega_{\rm sp})\cdot{\bf x}_\|}\,e^{-\beta_3(k_0,\,\omega_{\rm sp})x_3}\ ,
\label{e58}
\end{equation}
where ${\bf x}_\|=\{x_1,\,x_2\}$, $\hat{\bf k}_0$ and $\hat{\bf x}_3$ are
the unit vectors in the ${\bf k}_0=k_0(\omega_{\rm sp})\{\cos\theta_0,\,\sin\theta_0\}$
and $x_3$ directions, $E_{\rm sp}$ is the field amplitude, $\omega_{\rm sp}$ is
the field frequency, $\theta_0$ is the angle of the incident surface plasmon-polariton
wave with respect to the $x_1$ direction, ${\bf D}_0=\{-x_{1g},\,-x_{2g}\}$ is the
position vector of the center of a surface grating, and the two wave numbers in Eq.\,(\ref{e58}) are given by

\begin{equation}
k_0(\omega_{\rm sp})=\frac{\omega_{\rm sp}}{c}\sqrt{
\frac{\epsilon_{\rm d}\,\epsilon_{\rm M}(\omega_{\rm sp})}{
\epsilon_{\rm d}+\epsilon_{\rm M}(\omega_{\rm sp})}}\ ,
\label{e59}
\end{equation}
and

\begin{equation}
\beta_3(k_0,\omega_{\rm sp})=\sqrt{k^2_0(\omega_{\rm sp})
-\frac{\omega_{\rm sp}^2}{c^2}}
\label{e60}
\end{equation}
with ${\rm Re}[k_0(\omega_{\rm sp})]\geq 0$ and ${\rm Re}[\beta_3(k_0,\omega_{\rm sp})]\geq 0$.
Here, the in-plane wave vector $k_0$ is produced by the surface-grating diffraction
of the $p$-polarized normally-incident light, which in turn determines the resonant
frequency $\omega_{\rm sp}$ of the surface plasmon-polariton mode.
\medskip

In order for us to  explicitly determine the electric field dependence for
$\mbox{\boldmath${\cal P}$}^{\rm loc}({\bf r};\,\omega)$, we now turn
to the  quantum kinetics of electrons in a quantum dot. Here, the
optical polarization field $\mbox{\boldmath${\cal P}$}^{\rm loc}({\bf r};\,\omega)$
plays a unique role in bridging the gap between the classical Maxwell's
equations for electromagnetic fields and the quantum-mechanical
Schr\"odinger equation for electrons. The quantum kinetics of electrons in
photo-excited quantum dots should be adequately described  by the so-called
semiconductor Bloch equations\,\cite{r3,r4,r5}  which are a generalization
of the well-known optical Bloch equations in two ways, namely the incorporation
of electron scattering by impurities, phonons and other electrons, as well as
the many-body effects on dephasing in the photo-induced optical coherence.
\medskip

For photo-excited spin-degenerate  electrons (holes) in the conduction (valence)
band, their semiconductor Bloch equations with $\ell(j)=1,\,2,\,\cdots$ are
given, within the rotating-wave approximation, by

\begin{equation}
\frac{dn^{\rm e(h)}_{\ell(j)}}{dt}=\frac{2}{\hbar}\,\sum\limits_{j(\ell)}\,{\rm Im}\left[\left(Y_\ell^j\right)^\ast\left({\cal M}^{\rm eh}_{\ell,j}-Y_\ell^j\,V^{\rm
eh}_{\ell,j;j,\ell}\right)\right]
+\left.\frac{\partial n^{\rm e(h)}_{\ell(j)}}{\partial t}\right|_{\rm rel}
-\delta_{\ell(j),1}\,{\cal R}_{\rm sp}\,n_1^{\rm e}\,n_1^{\rm h}\ ,
\label{e19}
\end{equation}
where ${\cal R}_{\rm sp}$ is the spontaneous emission rate, which should be calculated
by using the Kubo-Martin-Schwinger relation\,\cite{r13} and including band gap energy and
interband dipole moment  renormalizations, and $n^{\rm e(h)}_{\ell(j)}$
represents the electron (hole) level population. In Eq.\,(\ref{e19}),  the Boltzmann-type
scattering term for non-radiative energy relaxation of electrons (holes) is
$\displaystyle{\left.\frac{\partial n^{\rm e(h)}_{\ell(j)}}{\partial t}\right|_{\rm rel}
={\cal W}_{\ell(j)}^{{\rm in}}(1-n_{\ell(j)}^{\rm e(h)})-
{\cal W}_{\ell(j)}^{{\rm out}}\,n_{\ell(j)}^{\rm e(h)}}$,
where ${\cal W}_{\ell(j)}^{{\rm in}}$ and ${\cal W}_{\ell(j)}^{{\rm out}}$
are the scattering-in and scattering-out rates for electrons (holes), respectively,
and should be calculated by including
carrier-carrier and carrier-(optical) phonon interactions.
Moreover, we know from Eq.\,(\ref{e19}) that the total number
$N_{\rm e(h)}(t)$ of photo-excited electrons (holes) is conserved.
\medskip

The induced optical polarization  in the semiconductor Bloch equations with
$\ell(j)=1,\,2,\,\cdots$ satisfies  the following equations for a spin-averaged
e-h plasma,  i.e.,

\[
i\hbar\,\frac{d}{dt}Y_\ell^j=\left[\overline{\varepsilon}^{\rm
e}_\ell(\omega\vert t)+\overline{\varepsilon}^{\rm
h}_j(\omega\vert t)-\hbar(\omega+i\gamma_0)\right]Y_\ell^j+\left(1-n^{\rm
e}_\ell-n^{\rm h}_j\right)\left({\cal M}^{\rm eh}_{\ell,j}-Y_\ell^j\,V^{\rm
eh}_{\ell,j;j,\ell}\right)
\]
\[
+Y_\ell^j\left[\sum\limits_{j_1}\,n^{\rm h}_{j_1}\left(V^{\rm hh}_{j,j_1;j_1,j}-V^{\rm hh}_{j,j_1;j,j_1}\right)-\sum\limits_{\ell_1}\,n^{\rm e}_{\ell_1}\,V^{\rm eh}_{\ell_1,j;j,\ell_1}\right]
\]
\begin{equation}
+Y_\ell^j\left[\sum\limits_{\ell_1}\,n^{\rm e}_{\ell_1}
\left(V^{\rm ee}_{\ell,\ell_1;\ell_1,\ell}-V^{\rm ee}_{\ell,\ell_1;
\ell,\ell_1}\right)-\sum\limits_{j_1}\,n^{\rm h}_{j_1}\,V^{\rm eh}_{\ell,j_1;j_1,\ell}\right]\ ,
\label{e22}
\end{equation}
where $Y^j_\ell$ represents the induced interband optical coherence,
$\hbar\gamma_0=\hbar\gamma_{\rm eh}+\hbar\gamma_{ext}$ is the energy level
broadening (due to finite carrier lifetime plus the radiation loss of an external
evanescent field), $\overline{\varepsilon}^{\rm e(h)}_{\ell(j)}(\omega\vert t)$ is
the kinetic energy of dressed single electrons (holes) (see supplementary materials). In Eq.\,(\ref{e22}), the
diagonal dephasing of $Y^j_\ell$, the renormalization of interband Rabi coupling, the
renormalization of electron and hole energies, as well as the
exciton binding energy, are all taken into account. Since the
e-h plasma is not spin-dependent, they may be excited by both left and right
circularly polarized light. The off-diagonal dephasing of $Y^j_\ell$ has
been neglected due to low carrier density in quantum dots.
In Eqs.(\ref{e19}) and (\ref{e22}), we have introduced the Coulomb matrix elements
$V^{\rm ee}_{\ell_1,\ell_2;\ell_3,\ell_4}$, $V^{\rm hh}_{j_1,j_2;j_3,j_4}$ and $V^{\rm eh}_{\ell,j;j^\prime,\ell^\prime}$, for electron-electron,
hole-hole and e-h interactions, respectively.
\medskip

The steady-state solution of Eq.\,(\ref{e22}), i.e., subject to  the condition that
$dY^j_\ell/dt=0$, has been obtained as

\begin{equation}
Y^j_\ell(t\vert\omega)=\left[\frac{1-n^{\rm
e}_\ell(t)-n_j^{\rm
h}(t)}{\hbar(\omega+i\gamma_0)-\hbar\overline{\Omega}^{\rm
eh}_{\ell,j}(\omega\vert t)}\right]{\cal M}^{\rm eh}_{\ell,j}(t)\ ,
\label{e23}
\end{equation}
where the photon and Coulomb renormalized interband energy level separation
$\hbar\overline{\Omega}^{\rm eh}_{\ell,j}(\omega\vert t)$ is given by

\[
\hbar\overline{\Omega}^{\rm eh}_{\ell,j}(\omega\vert t)=\overline{\varepsilon}^{\rm e}_\ell(\omega\vert t)+\overline{\varepsilon}^{\rm h}_j(\omega\vert t)-V^{eh}_{\ell,j;j,\ell}+\sum\limits_{\ell_1}\,n^{\rm e}_{\ell_1}(t)\left(V^{\rm ee}_{\ell,\ell_1;\ell_1,\ell}-V^{\rm ee}_{\ell,\ell_1;\ell,\ell_1}\right)
\]
\begin{equation}
+\sum\limits_{j_1}\,n^{\rm h}_{j_1}(t)\left(V^{\rm hh}_{j,j_1;j_1,j}-V^{\rm hh}_{j,j_1;j,j_1}\right)-\sum\limits_{\ell_1\neq \ell}\,n^{\rm e}_{\ell_1}(t)\,V^{\rm eh}_{\ell_1,j;j,\ell_1}-\sum\limits_{j_1\neq j}\,n^{\rm h}_{j_1}(t)\,V^{\rm eh}_{\ell,j_1;j_1,\ell}\ ,
\label{e24}
\end{equation}
and the matrix elements employed in Eqs.\,(\ref{e19}) and (\ref{e22}) for the
Rabi coupling between photo-excited carriers and an evanescent pump field
$\displaystyle{\mbox{\boldmath$E$}({\bf r};\,t)=\theta(t)\,\mbox{\boldmath$E$}({\bf r};\,
\omega)\,e^{-i\omega t}}$ are given by

\begin{equation}
{\cal M}^{\rm eh}_{\ell,j}(t)=-\delta_{\ell,1}\,\delta_{j,1}\,
\theta(t)\,\left[\mbox{\boldmath$E$}^{\rm eh}_{\ell,j}(\omega)
\cdot{\bf d}_{\rm c,v}\right]\ .
\label{e31}
\end{equation}
In this notation, $\theta(x)$ is the Heaviside unit step function, the
static interband dipole moment denoted by  ${\bf d}_{\rm c,v}$ is given by\,\cite{r14,r15} (see supplementary materials)

\begin{equation}
{\bf d}_{\rm c,v}=\int d^3{\bf r}\left[u_{\rm c}({\bf
r})\right]^\ast\,{\bf r}\,u_{\rm v}({\bf r})={\bf d}_{\rm c,v}^\ast\ ,
\label{e32}
\end{equation}
where $u_{\rm c}({\bf r})$ and $u_{\rm v}({\bf r})$ are the Bloch functions associated with conduction and valence bands at the $\Gamma$-point in the first Brillouin zone of the host semiconductor, and the effective electric field coupled to the quantum dot is evaluated by

\begin{equation}
\mbox{\boldmath$E$}^{\rm eh}_{\ell,j}(\omega)=\int d^3{\bf r}\left[\psi^{\rm e}_\ell({\bf r})\right]^\ast\mbox{\boldmath$E$}({\bf r};\,\omega)\left[\psi^{\rm h}_j({\bf r})\right]^\ast\ ,
\label{e33}
\end{equation}
where $\psi^{\rm e(h)}_{\ell(j)}({\bf r})$ is the envelope function of electrons (holes) in the quantum dot.
Next, the photo-induced interband optical polarization
$\mbox{\boldmath${\cal P}$}^{\rm loc}({\bf r};\,\omega)$ by dressed
electrons in the quantum dot is given by\,\cite{r2}

\[
\mbox{\boldmath${\cal P}$}^{\rm loc}({\bf r};\,\omega)=2\left|\xi_{\rm QD}({\bf
r})\right|^2\,{\bf d}_{\rm c,v}\left\{\int d^3{\bf r}^\prime\,\psi^{\rm e}_1({\bf
r}^\prime)\,\psi^{\rm h}_1({\bf r}^\prime)\right\}
\]
\begin{equation}
\times\,\frac{1}{\hbar}\,\lim\limits_{t\to\infty}\left[\frac{1-n^{\rm e}_1(t)-n_1^{\rm
h}(t)}{\omega+i\gamma_0-\overline{\Omega}^{\rm
eh}_{1,1}(\omega\vert t)}\right]{\cal M}^{\rm eh}_{1,1}(t)\ , \label{e45}
\end{equation}
where the profile function $|\xi_{\rm QD}({\bf r})|^2$ comes from the quantum confinement inside a quantum dot.
\medskip

In our numerical calculations, we chose the quantum dot dimensions as
$210$\,\AA\ and $100$\,\AA\ for  along the $x$ and $y$ directions, respectively,
$m^\ast_{\rm e}=0.067\,m_0$ and $m^\ast_{\rm h}=0.62\,m_0$ for the electron and
hole effective masses, in terms of the free electron mass $m_0$,
$\theta_0=45^{\rm o}$, $x_{1g}=x_{2g}=610$\,\AA, $\epsilon_b=12$ for
the quantum dot, $\epsilon_d=12$ for the cladding layer,
$\epsilon_s=11$ and $\epsilon_{\infty}=13$ for the static and optical dielectric
constants, $\hbar\Omega_0=36$\,meV for the energy of optical phonons,
$\hbar\Gamma_{\rm ph}=3$\,meV ($=\hbar\gamma_0$) for the phonon broadening,
$z_0=610$\,\AA, and $T=300$\,K for the lattice temperature.
The silver plasma frequency is $13.8\times10^{15}$\,Hz and the silver plasma dephasing
parameter is $0.1075\times10^{15}$\,Hz. The energy gap $E_{\rm G}$ for the
active quantum-dot material is $1.927$\,eV at $T=300$\,K.
\medskip

\begin{figure}[p]
\centering
\epsfig{file=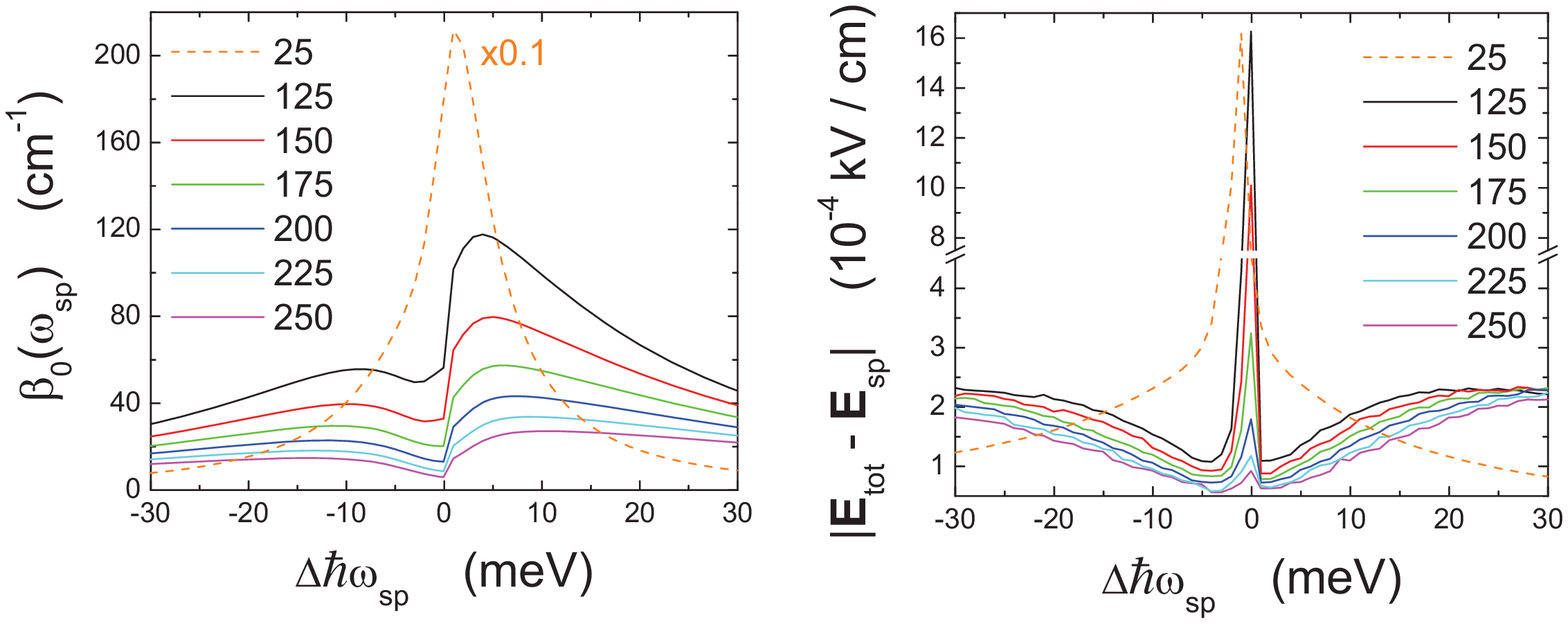,width=0.85\textwidth}
\epsfig{file=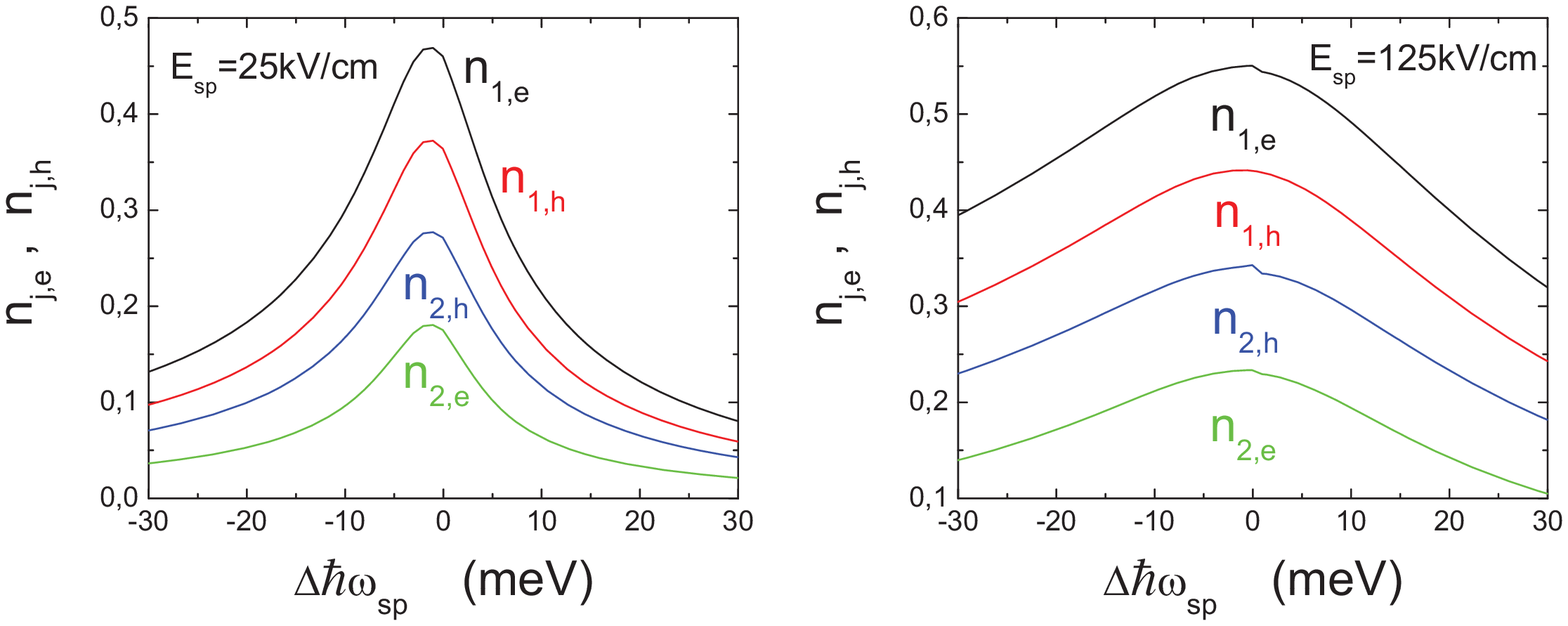,width=0.85\textwidth}
\caption{\label{f1}
(Color online) Absorption coefficients $\beta_0(\omega_{\rm sp})$
(upper-left) and scattering field $|{\bf E}_{\rm tot}-{\bf E}_{\rm  sp}|$
at the quantum dot (upper-right), as well as the energy-level
occupations for electrons $n_{\ell,e}$ and holes $n_{j,h}$ (lower)
as functions of the frequency detuning
$\Delta\hbar\omega_{\rm sp}$ (see text).
The results for various amplitudes $E_{\rm sp}$ of an SPP wave with frequency
$\omega_{\rm sp}$ are presented in the upper panels, along with a comparison of the
energy-level occupations for
$E_{\rm sp}=25$ and
$125$\,kV/cm in the lower panels. The label $\times 0.1$ in the upper
panel indicates that the result is multiplied by a factor of $0.1$.}
\end{figure}

Figure\ \ref{f1} presents the absorption coefficient $\beta_0(\omega_{\rm sp})$ for an
SPP wave by a quantum dot\,\cite{r13}, the scattering field $|{\bf E}_{\rm tot}-{\bf E}_{\rm sp}|$
of the SPP wave, and the energy-level occupations for electrons $n_{\ell,e}$ and
holes $n_{j,h}$ with $\ell,\,j=1,\,2$ as functions of the frequency detuning
$\Delta\hbar\omega_{\rm sp}\equiv\hbar\omega_{\rm sp}-(E_{\rm G}+\varepsilon_{1,e}
+\varepsilon_{1,h})$ with bare energies $\varepsilon_{1,e}$, $\varepsilon_{1,h}$
for ground-state electrons and holes. A dip is observed at resonance $\Delta
\hbar\omega_{\rm sp}=0$ in the upper-left panel, which becomes deeper with decreasing
amplitude $E_{\rm sp}$ of the SPP wave in the strong-coupling regime due to a reduction
of saturated absorption.  However, this dip disappears when $E_{\rm sp}$ drops to
$25$\,kV/cm in the weak-coupling limit due to the suppression of the photon dressing effect,
which is accompanied by a one-order of magnitude increase in the absorption-peak strength.
The dip in the upper-left panel corresponds to a peak in the scattering field, as may
be seen from the upper-right panel of this figure. The scattering field increases with
the frequency detuning away from resonance, corresponding to decreasing absorption.
Consequently, two local minima appear on both sides of resonance for the scattering field in
the strong-coupling regime. Maxwell-Bloch equations couple the field dynamics outside
a quantum dot with the electron dynamics inside the dot. At $E_{\rm sp}=125$\,kV/cm
in the lower-right-hand panel, we obtain peaks in energy-level occupations at resonance, which
are broadened by the finite carrier lifetime as well as the optical power of the SPP wave.
Moreover, jumps in the energy level occupation  may be seen at resonance due to Rabi
splitting of the energy levels in the dressed electron states. The effect of resonant phonon
absorption also plays a significant role in the finite value of $n_{2,e}$ with energy-level
separations $\varepsilon_{2,e}-\varepsilon_{1,e}\approx\hbar\Omega_0$. However, as $E_{\rm sp}$
decreases to $25$\,kV/cm in the lower-left panel, peaks in the energy-level occupations are
greatly sharpened and negatively shifted due to the suppression of the broadening from the
optical power and the excitonic effect, respectively. Additionally, jumps in the energy level
occupations become invisible because the Rabi-split energy gap in this case is much smaller
than the energy-level broadening from the finite lifetime of electrons
(i.e., substantially  dampened Rabi oscillations between the first electron and hole levels).
\medskip

\begin{figure}[p]
\centering
\epsfig{file=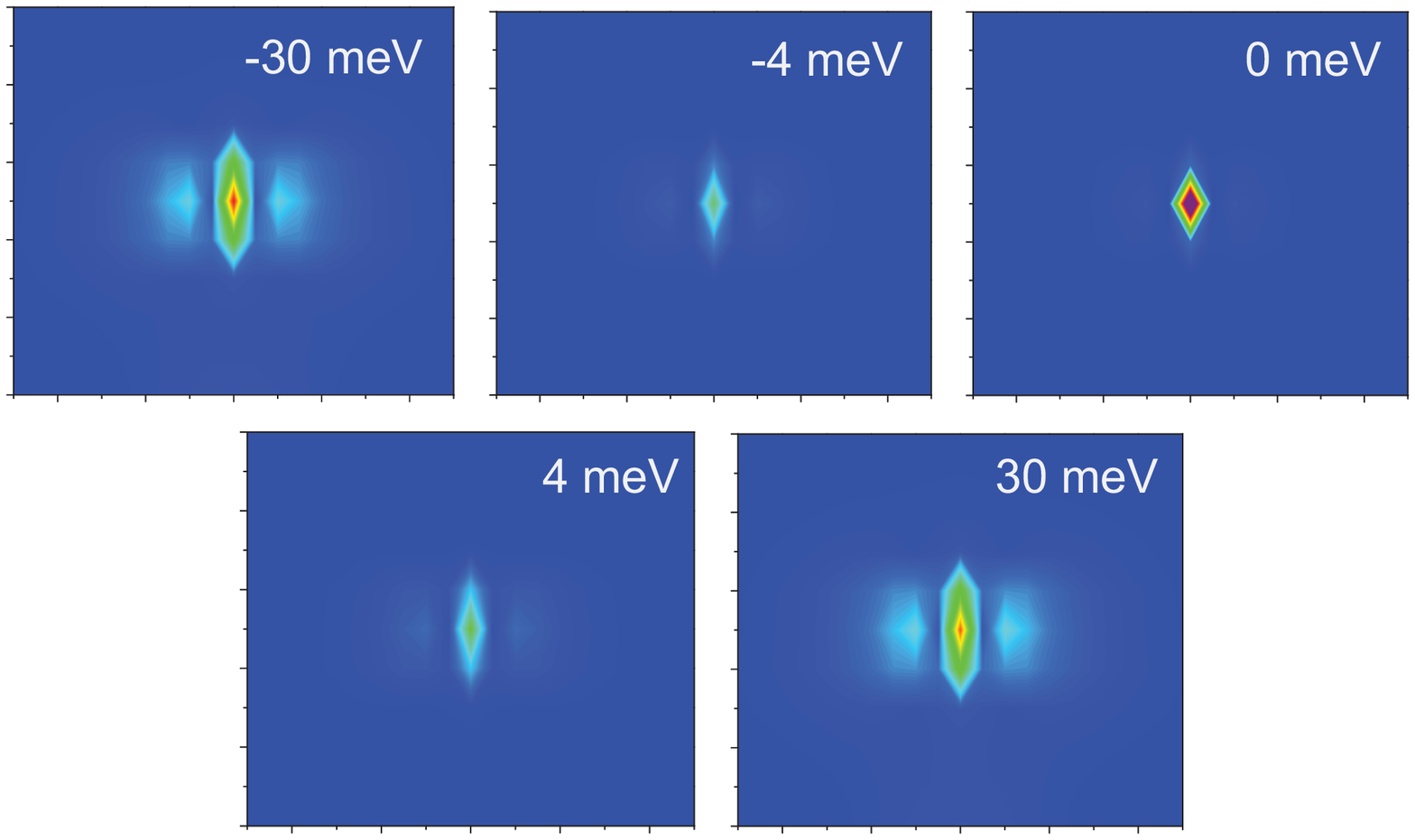,width=0.75\textwidth}
\vspace{1cm}
\qquad
\epsfig{file=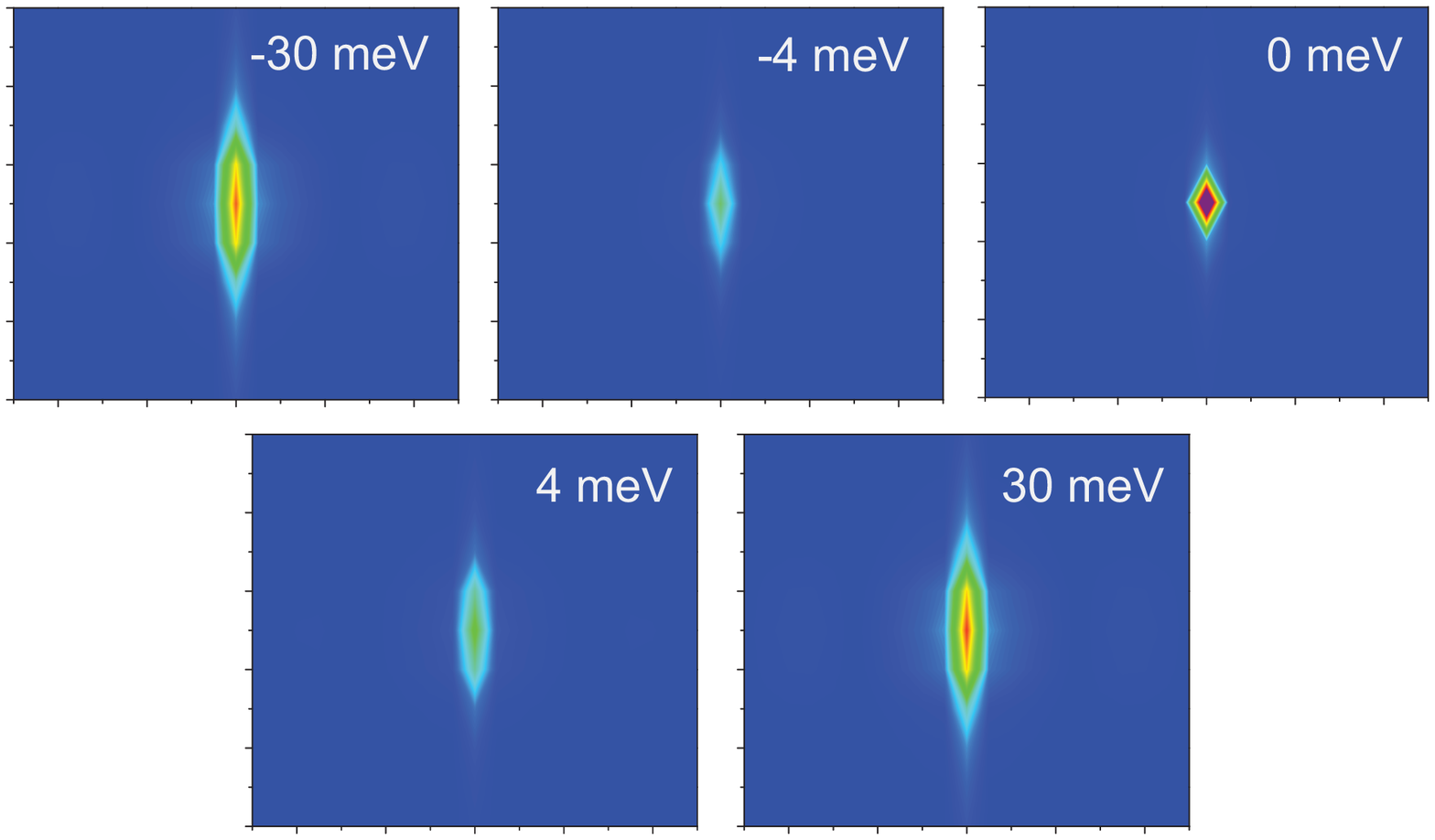,width=0.75\textwidth}
\caption{\label{f2}
(Color online) Scattering field color maps for $|E^\nu_{\rm tot}-E^\nu_{\rm sp}|/E_{\rm sp}$
(with $y=0$) around a quantum dot above a metallic surface in the $x$ ($\nu=1$, upper
panel) and $z$ ($\nu=3$, lower panel) directions, respectively,
with varying frequency detuning $\Delta\hbar\omega_{\rm sp}=-30$, $-4$, $0$, $4$ and $30$\,meV.
We chose $E_{\rm sp}=150$\,kV/cm. The color scales (blue-to-red)
for all values of $\Delta\hbar\omega_{\rm sp}$ are $0$--$1\times 10^{-7}$ in
the upper panel and $0$--$1.6\times 10^{-6}$ in the lower panel except for
$\Delta\hbar\omega_{\rm sp}=0$ where the color scales are $0$--$4.5\times 10^{-7}$
in the upper panel and $0$--$7\times 10^{-6}$ in the lower panel.
The use of different scales is to  clearly display spatial changes in
the intensity distribution.}
\end{figure}

Although the interband dipole moment of a quantum dot is isotropic in space,
the scattering field (see Fig.\,\ref{f2} with
$E_{\rm sp}=150$\,kV/cm) in the $x$ direction (upper panel) and
in the $z$ direction (lower panel) are still different
due to the presence of a metallic surface perpendicular to the $z$ direction
in our system. However, this isotropic intensity distribution is mostly recovered
at $\Delta\hbar\omega_{\rm sp}=0$. Specifically, the scattering field in the $z$ direction is one
order of magnitude larger than that in the $x$ direction. The field pattern in the
lower panel tends to spread in the $z$ direction, while the pattern in the upper
panel distributes in the $x$ direction. From this figure, we also find that the
intensities in both panels follow the pattern of strong-weak-strong-weak-strong as
the frequency detuning is swept across $\Delta\hbar\omega_{\rm sp}=0$,
which agrees with the observation of the scattering field at the quantum dot in
the upper-right panel of Fig.\,\ref{f1}.
\medskip

\begin{figure}[p]
\centering
\epsfig{file=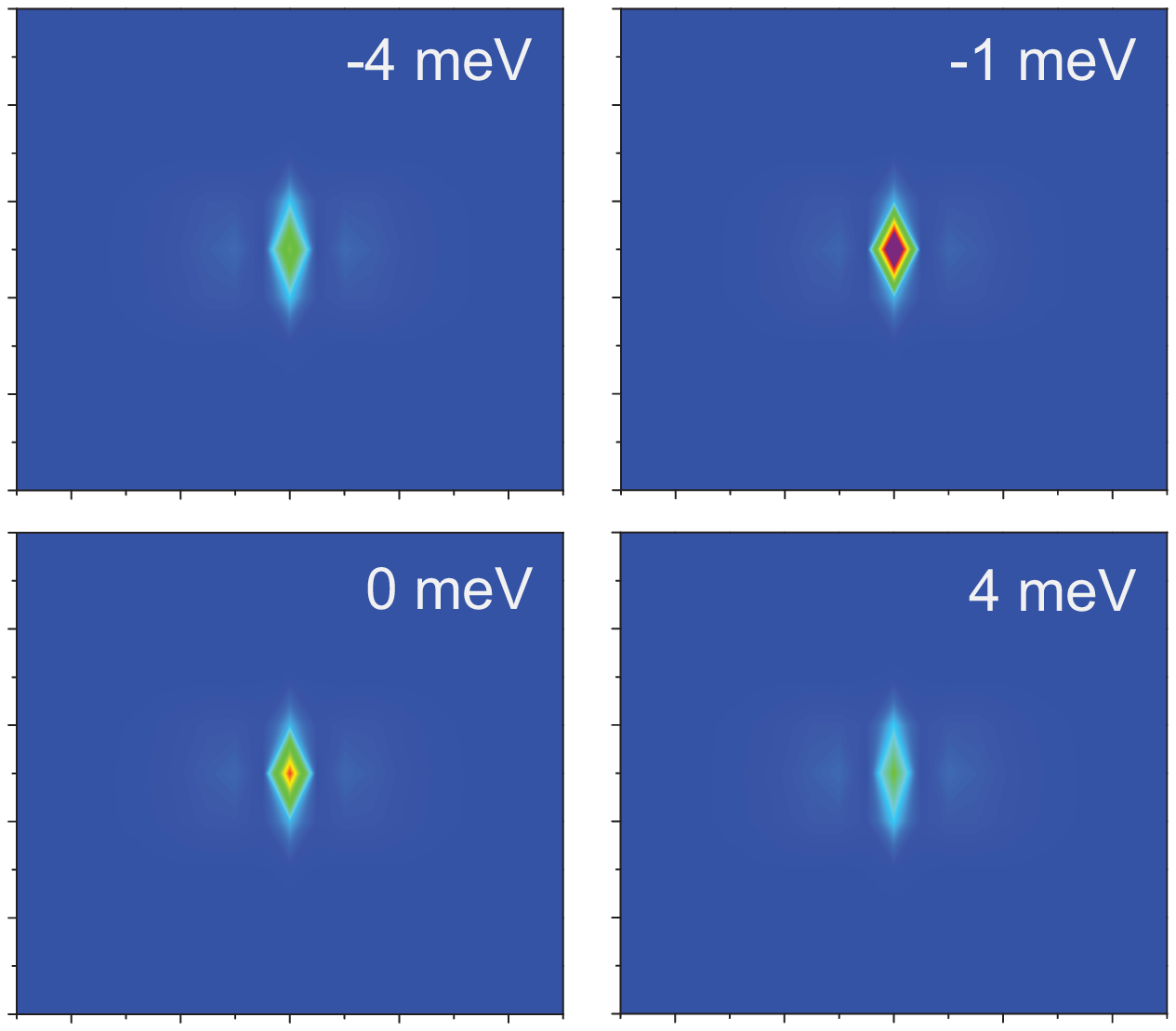,width=0.55\textwidth}
\vspace{1cm}
\qquad
\epsfig{file=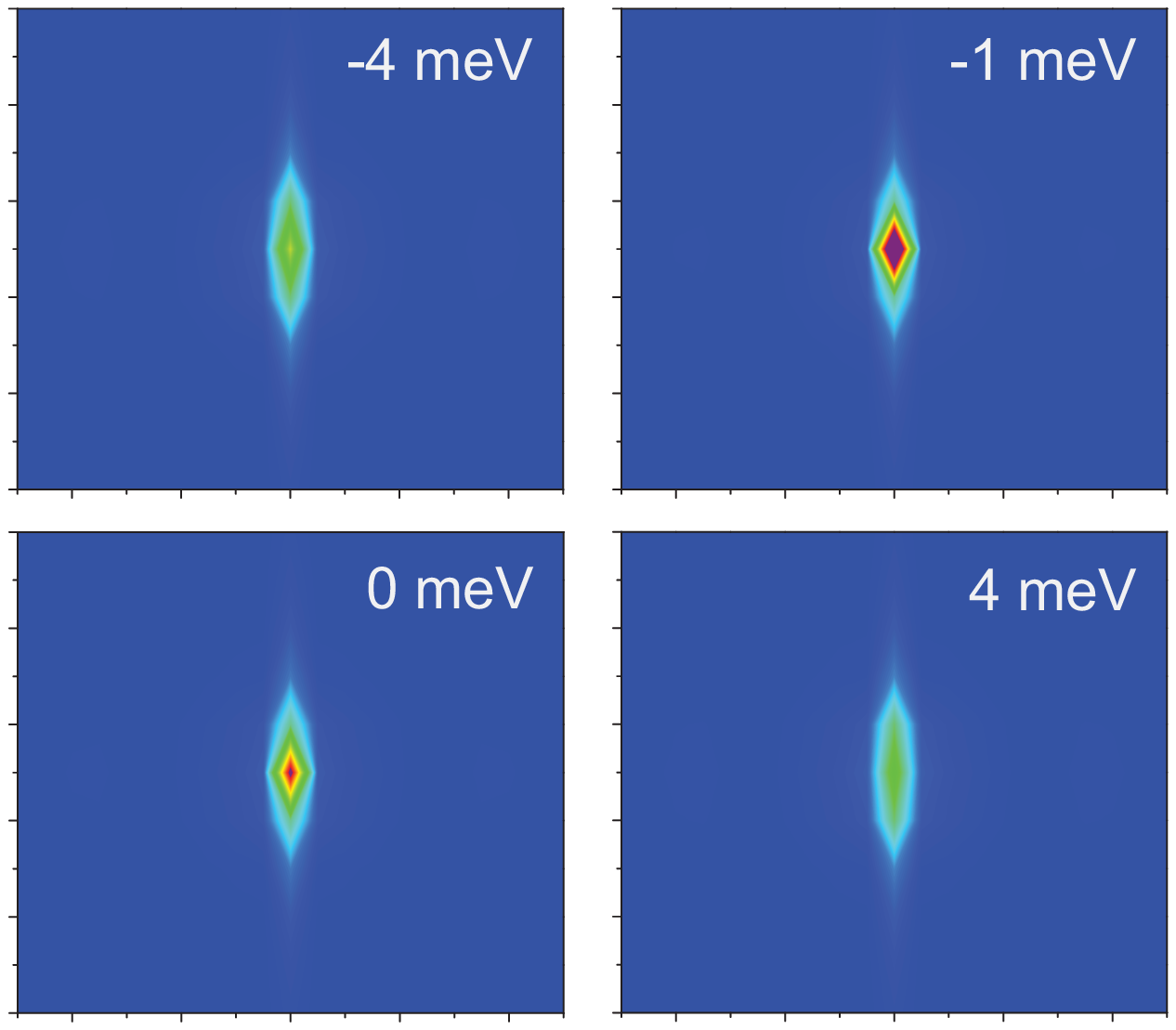,width=0.55\textwidth}
\caption{\label{f3}
(Color online) Scattering field color maps for
$|E^\nu_{\rm tot}-E^\nu_{\rm sp}|/E_{\rm sp}$
(with $y=0$) around a quantum dot above the metallic surface in the
$x$ ($\nu=1$, upper panel) and $z$ ($\nu=3$, lower panel) directions, respectively,
with varying frequency detuning $\Delta\hbar\omega_{\rm sp}=-4$, $-1$, $0$ and $4$\,meV.
Here, $E_{\rm sp}=25$\,kV/cm is assumed. The color scales (blue-to-red)
for all values of $\Delta\hbar\omega_{\rm sp}$ are $0$--$1.4\times 10^{-6}$ in the upper
panel and $0$--$2\times 10^{-5}$ in the lower panel except for
$\Delta\hbar\omega_{\rm sp}=-1$\,meV where the color scales
are $0$--$4\times 10^{-6}$ in the upper panel and $0$--$7\times 10^{-5}$ in the lower panel.
Again, the use of different scales is for a clear display of spatial changes in the intensity distribution.}
\end{figure}

Color maps for the scattering field around a quantum dot displayed in
Fig.\,\ref{f2} are for  strong coupling between the dot and an SPP wave.
We present in Fig.\,\ref{f3} the scattering field maps in the weak-coupling
regime, where the strong-weak-strong-weak-strong pattern in the strong-coupling
regimes has been changed to a weak-strong-weak pattern. Moreover, the SPP-wave
frequency for the resonant scattering field has been shifted from
$\Delta\hbar\omega_{\rm sp}=0$ to $\Delta\hbar\omega_{\rm sp}=1$\,meV, demonstrating a positive
depolarization shift of the optical excitation energy, as may be verified from the
upper left-hand panel of Fig.\,\ref{f1}. However, this depolarization effect is
completely masked by the occurrence of a local minimum at $E_{\rm sp}=150$\,kV/cm.
\medskip

In conclusion, for a strong SPP wave, we have demonstrated the unique effect
of its resonant scattering by a dynamical semiconductor quantum dot very close
to the metal/dielectric interface. We have also predicted correlation
between a resonant peak in the scattering field spectrum and a resonant local
 minimum in the absorption spectrum of the SPP wave.
\medskip

In this Letter, we only investigated the coupling between an SPP wave and a single quantum
dot for the simplest case. Our formalism may be generalized in a straightforward way to
include multiple quantum dots close to the surface of a metallic film. The open surface of the
metallic film provides an easy solution to perform biochemical and biomedical tests
under a microscope in a laboratory setting. If the quantum dots are further coated
with specially-selected  chemically-reactive molecules, they should be expected to
be able to adhere to special  target tissue if the chemical properties are matched with each other.
Therefore, the spontaneous emission by electrons in these quantum dots may be employed
non-invasively for near-field imaging of target tissue with very high brightness and
spatial resolution.
\medskip

Additionally, instead of coupling to the lowest pair of e-h energy levels, we may
choose the surface plasmon frequency for a resonant coupling to the next pair
of e-h levels. In this case,  optical pumping from the localized
surface plasmon field may transfer a population inversion from the excited
pair to the ground pair of e-h levels by exchanging thermal
energy with lattice phonons, leading to a possible lasing action. Such a surface
plasmon   based quantum-dot laser would have a beam size as small as a few nanometers
(not limited by its wavelength), which is expected to be very useful to spatially
selective illumination of individual molecules (neuron cells) in low-temperature
photo-excited chemical reactions (optogenetics and neuroscience).
\medskip

\begin{acknowledgments}
DH would like to acknowledge the support by the Air Force Office of Scientific Research (AFOSR).
This research was also supported by  contract \# FA 9453-13-1-0291 of
AFRL.
\end{acknowledgments}

\appendix{}

\section{ELECTRONIC STATES OF A QUANTUM DOT}

We have employed a box-type potential with hard walls for a quantum dot, which is given by

\begin{equation}
V({\bf r})=\left\{\begin{array}{ll}
0\ , & \mbox{$0\leq x_i\leq L_i$ for $i=1,\,2,\,3$}\\
\infty\ , & \mbox{others}
\end{array}\right.\ ,
\label{a1}
\end{equation}
where the position vector ${\bf r}=(x_1,x_2,x_3)$, $L_1$, $L_2$ and $L_3$ are the widths of the potential in the $x_1$, $x_2$ and $x_3$ directions, respectively. The Schr\"odinger equation for a single electron or hole in a quantum dot is written as

\begin{equation}
-\frac{\hbar^2}{2m^\ast}\left[\frac{\partial^2}{\partial x_1^2}+\frac{\partial^2}{\partial x_2^2}+\frac{\partial^2}{\partial x_3^2}+V({\bf r})\right]\psi({\bf r})=\varepsilon\,\psi({\bf r})\ ,
\label{a2}
\end{equation}
where the effective mass $m^\ast$ is $m^\ast_{\rm e}$ for electrons or $m^\ast_{\rm h}$ for holes. The eigenstate wave function associated with Eq.\,(\ref{a2}) is found to be

\begin{equation}
\psi_{n_1,n_2,n_3}({\bf r})=\sqrt{\frac{2}{L_1}}\,\sin\left[\left(\frac{n_1\pi}{L_1}\right)x_1\right]\sqrt{\frac{2}{L_2}}\,\sin\left[\left(\frac{n_2\pi}{L_2}\right)x_2\right]
\sqrt{\frac{2}{L_3}}\,\sin\left[\left(\frac{n_3\pi}{L_3}\right)x_3\right]\ ,
\label{a3}
\end{equation}
which is same for both electrons and holes, and the eigenstate energy associated with Eq.\,(\ref{a2}) is

\begin{equation}
\varepsilon_{n_1,n_2,n_3}=\frac{\hbar^2}{2m^\ast}\left[\left(\frac{n_1\pi}{L_1}\right)^2+\left(\frac{n_2\pi}{L_2}\right)^2+\left(\frac{n_3\pi}{L_3}\right)^2\right]\ ,
\label{a4}
\end{equation}
where the quantum numbers $n_1,\,n_2,\,n_3=1,\,2,\,\cdots$.
\medskip

By using the calculated bare energy levels in Eq.\,(\ref{a4}), the dressed electron ($\lambda^{\rm e}_\alpha$) and hole ($\lambda^{\rm h}_\alpha$) energy levels under the rotating wave approximation take the form of

\begin{equation}
\lambda^{\rm e}_\alpha(\omega\vert t)=\lambda^{\rm h}_\alpha(\omega\vert t)=
\left\{\begin{array}{cc}
\frac{1}{2}\left(\hbar\omega+\sqrt{[{\cal E}_{\rm G}(T)+\varepsilon_\alpha^{\rm e}+\varepsilon_\alpha^{\rm h}-\hbar\omega]^2+4|{\cal M}^{\rm eh}_{\alpha,\alpha}(t)|^2}\,\right) &
\\
\mbox{if $\hbar\omega\leq {\cal E}_{\rm G}(T)+\varepsilon_\alpha^{\rm e}+\varepsilon_\alpha^{\rm h}$}\\
\\
\frac{1}{2}\left(\hbar\omega-\sqrt{[{\cal E}_{\rm G}(T)+\varepsilon_\alpha^{\rm e}+\varepsilon_\alpha^{\rm h}-\hbar\omega]^2+4|{\cal M}^{\rm eh}_{\alpha,\alpha}(t)|^2}\,\right) &
\\
\mbox{if $\hbar\omega\geq {\cal E}_{\rm G}(T)+\varepsilon_\alpha^{\rm e}+\varepsilon_\alpha^{\rm h}$}
\end{array}\right.\ ,
\label{a5}
\end{equation}
where the composite index $\alpha=\{n_1,\,n_2,\,n_3\}$. Moreover, we get the energy levels of dressed electrons in the lab frame, i.e.,
$\overline{\varepsilon}^{\rm e}_{\alpha}(\omega\vert t)=\lambda^{\rm e}_\alpha(\omega\vert t)+(\varepsilon^{\rm e}_\alpha-\varepsilon^{\rm h}_\alpha)/2$ and $\overline{\varepsilon}^{\rm e}_{\ell}(\omega\vert t)=\varepsilon_{\ell}^{\rm e}+{\cal E}_{\rm G}(T)/2$ for $\ell\neq\alpha$. Similarly, we obtain the energy levels of dressed holes $\overline{\varepsilon}^{\rm h}_{\alpha}(\omega\vert t)=\lambda^{\rm h}_\alpha(\omega\vert t)+(\varepsilon^{\rm h}_\alpha-\varepsilon^{\rm e}_\alpha)/2$
and $\overline{\varepsilon}^{\rm e}_j(\omega\vert t)=\varepsilon_j^{\rm h}+{\cal E}_{\rm G}(T)/2$ for $j\neq\alpha$.
\medskip

The interband dipole moment ${\bf d}_{\rm c,v}=d_{\rm c,v}\,\hat{\bf e}_{\rm d}$ at the isotropic $\Gamma$-point can be calculated according to the Kane approximation

\begin{equation}
d_{\rm c,v}=\sqrt{\frac{e^2\hbar^2}{2m_0\,{\cal E}_{\rm
G}(T)}\left(\frac{m_0}{m^\ast_{\rm e}}-1\right)}\ . \label{a9}
\end{equation}
Furthermore, the direction of the dipole moment $\hat{\bf e}_{\rm d}$ is determined by the quantum-dot energy levels in resonance with the photon energy $\hbar\omega$.

\newpage

\end{document}